\pdfoutput=1

\documentclass[11pt]{article}

\usepackage{graphicx} 
\usepackage{fancybox} 
\usepackage{indentfirst} 
\usepackage{amsthm} 
\usepackage{latexsym} 
\usepackage{amsmath} 
\usepackage{amssymb} 
\usepackage{amsbsy} 
\usepackage{array} 
\usepackage{enumitem} 
\usepackage{subfiles} 
\usepackage{titlesec} 
\usepackage{titletoc}
\usepackage{chngcntr} 
\usepackage{pdflscape} 
\usepackage{afterpage}
\usepackage[ruled,vlined]{algorithm2e}  
\usepackage{capt-of} 
\usepackage{multirow} 
\usepackage{fancyhdr} 

\usepackage{listings}
\usepackage{float}
\usepackage{subcaption}

\usepackage[nonumberlist, nopostdot, nogroupskip, acronym]{glossaries}
\usepackage{glossary-superragged}
\setglossarystyle{superraggedheaderborder}
\usepackage{setspace}
\usepackage{parskip}

\usepackage{amscd,amsfonts,enumerate}
\usepackage{color, fancyhdr, wrapfig}
\usepackage{hyperref}
\usepackage{longfbox}
\usepackage{tcolorbox}
\usepackage{float}

\usepackage[utf8]{inputenc}
\usepackage{varwidth}
\usepackage[thinlines]{easytable}
\usepackage{tabularx}
\usepackage{supertabular}
\usepackage{tikz}

\usepackage{booktabs}
\usepackage{caption}
\usepackage{graphics}
\usepackage{enumitem}

\usepackage{tocbasic}
\usepackage{enumerate}

\usepackage{blindtext}

\usepackage[utf8]{inputenc} 
\usepackage[T1]{fontenc}    
\usepackage{hyperref}       
\usepackage{url}            
\usepackage{booktabs}       
\usepackage{amsfonts}       
\usepackage{nicefrac}       
\usepackage{microtype}      
\usepackage{xcolor}         
\usepackage{listings}
\usepackage[export]{adjustbox}
\usepackage{times}
\usepackage{latexsym}

\usepackage[T1]{fontenc}

\usepackage[utf8]{inputenc}

\usepackage{microtype}

\usepackage{inconsolata}
\usepackage[]{acl}

\usepackage{times}
\usepackage{latexsym}

\usepackage[T1]{fontenc}

\usepackage[utf8]{inputenc}

\usepackage{microtype}

\usepackage{inconsolata}

\def\methodname{DocChecker }

\title{DocChecker: Bootstrapping Code Large Language Model for Detecting and Resolving Code-Comment Inconsistencies}

\author{Anh T. V. Dau \\
  FPT Software AI Center  \\
  Vietnam \\
  \texttt{anhdtv7@fpt.com} \\\And
  Jin L.C. Guo \\
  McGill University \\
  Canada \\
  \texttt{jguo@cs.mcgill.ca} \\\And
  Nghi D. Q. Bui \\
  Fulbright University \\
  Vietnam \\
  \texttt{nghi.bui@fulbright.edu.vn} \\}

\begin{document}
\maketitle
\begin{abstract}


Comments within source code are essential for developers to comprehend the code's purpose and ensure its correct usage. However, as codebases evolve, maintaining an accurate alignment between the comments and the code becomes increasingly challenging.
Recognizing the growing interest in automated solutions for detecting and correcting differences between code and its accompanying comments, current methods rely primarily on heuristic rules. In contrast, this paper presents DocChecker, a tool powered by deep learning. DocChecker is adept at identifying inconsistencies between code and comments, and it can also generate synthetic comments. This capability enables the tool to detect and correct instances where comments do not accurately reflect their corresponding code segments.
We demonstrate the effectiveness of DocChecker using the Just-In-Time and CodeXGlue datasets in different settings. Particularly, DocChecker achieves a new State-of-the-art result of 72.3\% accuracy on the Inconsistency Code-Comment Detection (ICCD) task and 33.64 BLEU-4 on the code summarization task against other Large Language Models (LLMs), even surpassing GPT 3.5 and CodeLlama.

DocChecker is accessible for use and evaluation. It can be found on \href{https://github.com/FSoft-AI4Code/DocChecker}{GitHub} and as an \href{http://4.193.50.237:5000/}{Online Tool}. For a more comprehensive understanding of its functionality, a demonstration video is available on \href{https://youtu.be/FqnPmd531xw}{YouTube}.
\end{abstract}

\section{Introduction}

\begin{figure}[t]
\centering
\small
\includegraphics[width=0.5\textwidth]{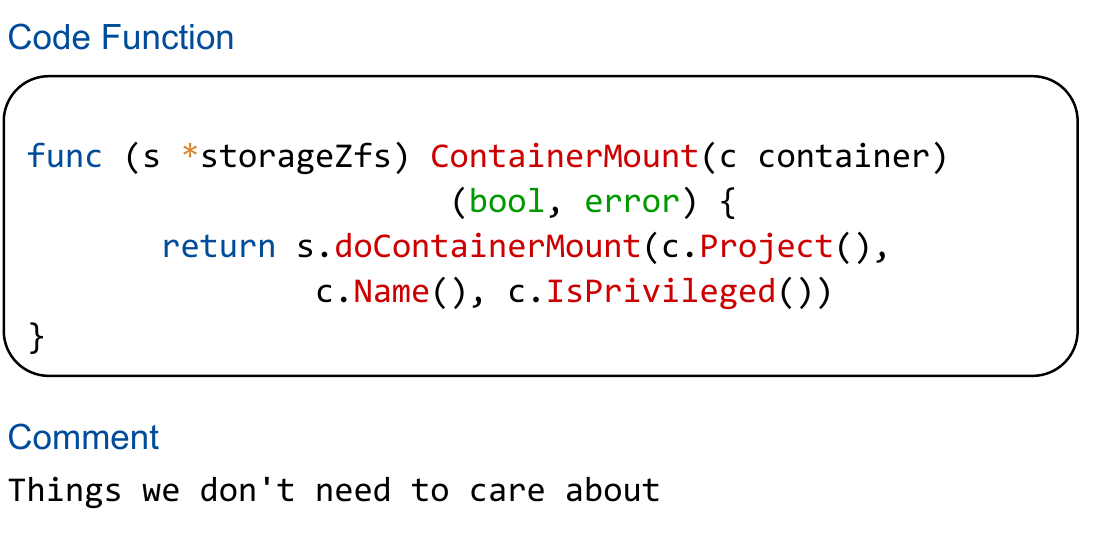}
\caption{An example of code-comment inconsistency from the CodeSearchNet dataset.}
\label{fig:inconsistent_sample}
\end{figure}

In the realm of Software Engineering, a critical challenge is the detection and resolution of inconsistencies between source code and its accompanying comments. Such inconsistencies, which may arise due to code changes not being reflected in comments or from initially inaccurate descriptions, can lead to significant issues in both understanding and maintaining software. An illustrative example of this inconsistency, sourced from the CodeSearchNet dataset, is depicted in Figure \ref{fig:inconsistent_sample} \cite{husain2019codesearchnet}. These disparities can cause software defects, degrade software quality, and lower developer productivity, as highlighted in recent studies \cite{wen2019large,6200082,panthaplackel2021deep,steiner2022code}. Moreover, the prevalent issue of code-comment conflicts in widely used datasets impacts the efficacy of code language models trained on them \cite{sun2022importance,shi2022we,manh2023vault}.

The advancements in Artificial Intelligence for Software Engineering (AI4SE) and the development of sophisticated code language models present an opportunity to address these challenges. These models, trained on extensive datasets that include both source code and natural language comments, have shown potential in understanding and processing code at an unprecedented scale \cite{wang2021codet5,nijkamp2022codegen,wang2023codet5+,codesearchsurvey}. However, the efficacy of these models is contingent upon the quality of their training data, underscoring the need for accurate and consistent code-comment pairs.

In this context, our work introduces DocChecker, a deep learning-based tool specifically designed for the task of Inconsistency Code-Comment Detection (ICCD). Leveraging the capabilities of AI4SE and the insights gained from advancements in code LLMs, DocChecker addresses the critical need for high-quality, consistent documentation in software development. The key idea is to leverage an encoder-decoder framework, pre-trained on code-text pairs. This pre-training employs a multi-faceted approach, including contrastive learning to \textit{bootstrap} code and text features, binary classification to discern consistent from inconsistent pairs, and text generation to create coherent comments. The backbone of this system is the UniXcoder~\cite{guo2022unixcoder}, chosen for its effectiveness and efficiency in handling multi-modal content. DocChecker is specifically designed not only to detect but also to resolve inconsistencies between code and comments by generating replacement comments that accurately reflect the current state of the codebase. Furthermore, compared to state-of-the-art CodeLLMs, 
our method excels significantly on ICCD and code summarization tasks. \methodname outperforms StarCoder by 30\% and surpasses GPT-3.5 and CodeLlama by 10\% in terms of accuracy, even though such models are pre-trained on larger-scale datasets. In summary, the key contributions of DocChecker are:
\begin{itemize}[leftmargin=*]
 \item  We propose DocChecker, a tool built on a code language model, jointly pre-trained with three objectives: contrastive learning, binary classification, and text generation.

\item To the best of our knowledge, \methodname is the first AI4SE tool designed to identify and fix inconsistent code-text pairs.

\item The experiments show that \methodname attains state-of-the-art results on ICCD and code summarization, compared to existing methods and LLMs such as StarCoder, GPT-3.5, and CodeLlama.

\end{itemize}

\section{Related Work}
    \subsection{The Pre-trained Code Language Models}

In recent years, language modeling has made significant strides, particularly with the introduction of pre-trained Transformers\cite{vaswani2017attention,DBLP:conf/naacl/DevlinCLT19,radford2018improving}. Pioneered by CodeX \citet{codeX}, LLMs have demonstrated remarkable success in code processing, giving rise to code models such as StarCoder \citet{li2023starcoder}, CodeLlama \citet{roziere2023code}. CuBERT \citet{cubert} and CodeBERT \citet{feng2020codebert} were the first encoder-only models. CuBERT utilizes masked language modeling and next-sentence prediction objectives to pre-train on the corpus of Python, while CodeBERT is pre-trained on six programming languages using the replace token detection task. Although encoder-only models achieve high performance for code-understanding tasks, they need an additional decoder for tasks requiring generation. On the other hand, \citet{IntelliCode} and \citet{codeX} employ GPT for the code completion task, but they need to be more suboptimal for understanding tasks. Some others \citet{wang2021codet5,guo2022unixcoder,wang2023codet5+,nijkamp2022codegen} utilize the encoder-decoder framework to support both understanding and generation tasks. 
In this work, we propose DocChecker, which learns semantic relationships between code and text to detect inconsistent code-comment pairs and generate improved summary sentences for the replacement. 

\subsection{Code-Comment Inconsistency}

The source code comment is very important in understanding the meaning of the code function. The significance of comments aligning with the source code is divided into two categories: inconsistent code-comment detection and comment updates. 
\citet{detecting} measures the similarity between code functions and comments, identifying inconsistency when the score falls below a set threshold. 
\citet{panthaplackel2021deep} develops a deep learning-based approach to comprehend and establish relationships between comments and code changes.
Instead of using machine learning approaches, others propose rule-based methods for analysis.
\citet{ratol2017detecting} introduces Fraco, an Eclipse plugin for fragile comment detection during identifier renaming, while \citet{shi2022we} develops an automated code-comment cleaning tool for accurate noise detection in the CodeSearchNet dataset \citet{husain2019codesearchnet}.
Although rule-based methods are clear and straightforward, they struggle with new datasets and lack semantic understanding. 
Recent research explores automatic comment updating, with tools like CUP \cite{CUP} and HebCUP \cite{hebCUP} effective for simple changes (a single token change) but not for complex ones.
In contrast, our tool excels at detecting and updating inconsistent code-comment pairs.

\section{Overview of DocChecker}
In this section, we describe the \methodname tool as a Python package and demonstrate its user interface. For full customization and detailed documentation of DocChecker, users can reference our \href{https://github.com/FSoft-AI4Code/DocChecker}{GitHub} repository.

\subsection{Python Package}
\label{usage:package}
We bundle DocChecker into an easy-to-use library that can be installed via \href{https://pypi.org/project/docchecker/}{Pypi}. 

\begin{figure}[]
\centering
\small
\includegraphics[width=0.5\textwidth]{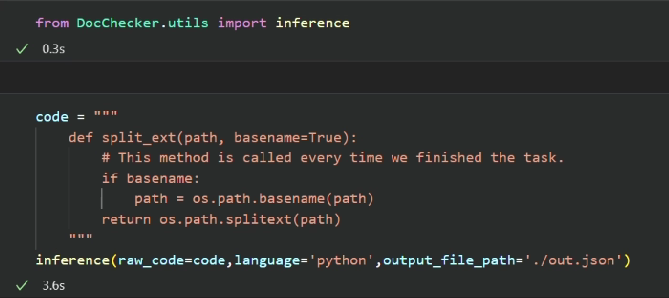}
\caption{Screenshot for the Input Example.
}
\label{fig:screenshot_input}
\end{figure}

\paragraph{Input:} The user must provide their source code file as well as the corresponding programming language. \methodname is able to extract all the code function and their metadata (e.g. function name) by using Tree-sister \footnote{https://github.com/tree-sitter/tree-sitter} parser. It supports 10 popular programming languages: Java, JavaScript, Python, Ruby, Rust, Golang, C\#, C++, C, and PHP. An example of how to use \methodname is illustrated in Figure \ref{fig:screenshot_input}.

\begin{figure}[]
\centering
\small
\includegraphics[width=0.5\textwidth]{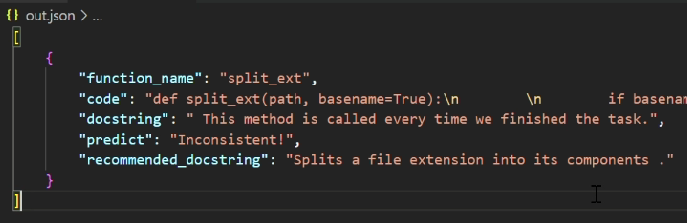}
\caption{Screenshot for the Output file Example.
}
\label{fig:screenshot_output}
\end{figure}

\paragraph{Output:} \methodname returns in the form of a comprehensive list of dictionaries corresponding in number to the received code functions, including the name of each function in the raw code, code snippet, associated docstring, as well as its prediction, and the recommended docstring. If a code-text pair is considered as \textit{“Inconsistent!”}, DocChecker will generate a complete docstring to replace the old ones; otherwise, it will keep the original version. Figure \ref{fig:screenshot_output} is a screenshot that shows the result of DocChecker's prediction.


\subsection{User Interface}
\label{usage:UI}
\begin{figure}[]
\centering
\small
\includegraphics[width=0.47\textwidth]{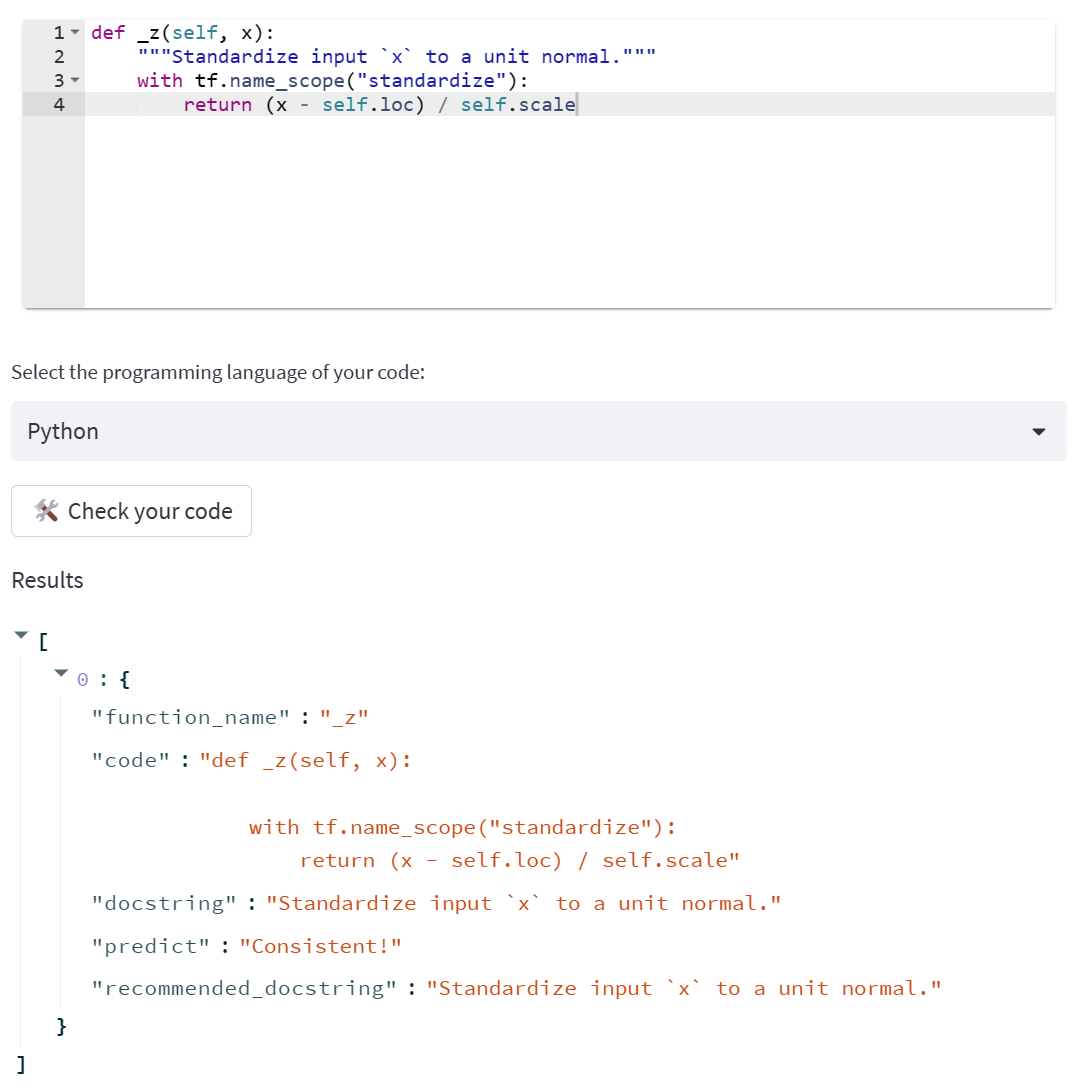}
\caption{Screenshot for the user interface.
}
\label{fig:screenshot_infer}
\end{figure}

We show a demo interface of DocChecker as depicted in Figure ~\ref{fig:screenshot_infer}. It consists of a coding field for directly entering source code or uploading existing code files, a select widget for specifying the programming language used for their code, and a button that serves as the trigger for starting the query process. When the front-end receives the query result, it displays the previously mentioned list of dictionaries. 
    
\section{Building Blocks of DocChecker}
    DocChecker is a neural architecture designed to learn from code-text pairs, leveraging an encoder-decoder model. This section outlines the architecture of DocChecker (see Section \ref{method:pretrain}), the objectives guiding its pre-training (Section \ref{method_objective}), and the specific setup used during pre-training (Section \ref{method:pretrain_setup}). Initially, the model undergoes pre-training focusing on contrastive learning and code-to-text generation, followed by fine-tuning for the specific Inconsistency Code-Comment Detection (ICCD) task.

\subsection{Architecture}
\label{method:pretrain}

\begin{figure*}[]
\centering
\small
\includegraphics[width=\textwidth]{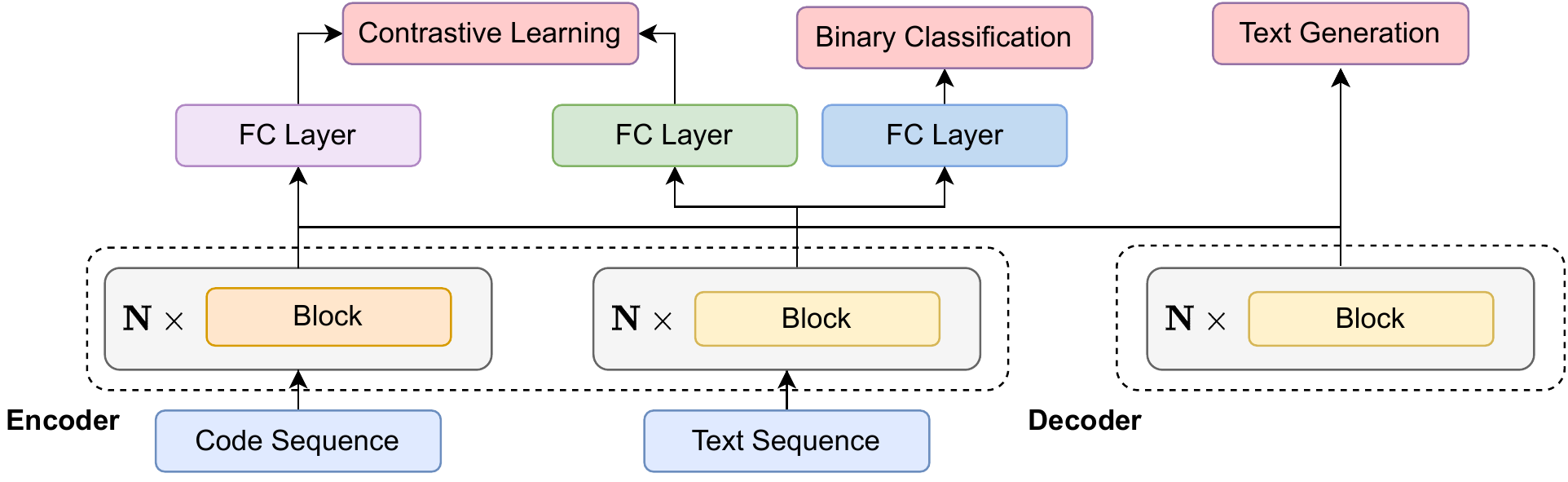}
\caption{Overview of the DocChecker framework.}
\label{fig:overview}
\end{figure*}

DocChecker's design is influenced by the effectiveness of pre-trained models. Instead of building from scratch, it utilizes existing pre-trained encoder-decoder models. For this project, we selected UniXcoder, an encoder-decoder model \cite{guo2022unixcoder}, as our backbone network due to its customizable nature and efficient performance with relatively fewer parameters (details in Section \ref{effective_backbone}).

\subsection{Pre-training Objectives}
\label{method_objective}
DocChecker's pre-training involves three primary objectives:
\paragraph{Code-Text Contrastive Learning (CTC):} This aims to align the feature spaces of the code and text encoders. By emphasizing similarities in positive code-text pairs and differentiating them from negative pairs, we enhance model accuracy. Negative samples are generated following the methodology in \cite{li2021align}, focusing on hard negative pairs based on contrastive similarity.

\paragraph{Binary Classification (BC):} This objective assesses the alignment between code and text. The model distinguishes between consistent (positive) and inconsistent (negative) code-text pairs, enhancing its ability to detect inconsistencies.

\paragraph{Text Generation (TG):} This focuses on creating a summary sentence from a given code snippet. By training the model to optimize cross-entropy loss in an autoregressive manner, it enhances the model's capacity to generate coherent comments.

In addition to these objectives, DocChecker benefits from multi-task learning, sharing the weights between the text encoder and decoder to improve text representation. Separate, fully-connected layers are utilized to capture task-specific differences and minimize task interference.

\subsection{Pre-training Setup}
\label{method:pretrain_setup}
DocChecker employs UniXcoder \cite{guo2022unixcoder}, renowned for its performance in multi-modal contexts and its unified cross-modal model. With 12 hidden layers, 768 hidden sizes, and 3072 intermediate sizes, UniXcoder's architecture of 124M parameters proves ideal for our purposes.

The pre-training dataset comes from the CodeXGLUE dataset \cite{lu2021codexglue}, encompassing six programming languages - Python, Javascript, Ruby, Go, Java, and PHP — sourced from open-source libraries. This diverse dataset is instrumental in ensuring the model's proficiency across a wide range of software engineering scenarios.

\section{Experiment Setup}
    
In this section, we first present the tasks and the datasets used to assess the performance of DocChecker. Then, we describe the baselines and metrics used for evaluation. 

\subsection{Tasks} 
\methodname is evaluated for two tasks: ICCD and Code Summarization.

\paragraph{ICCD:} Our task is given a comment \textit{$C$} with a corresponding code method \textit{$M$}, determine whether the comment \textit{$C$} is semantically out of sync with the code function \textit{$M$}. 
To address this challenge, we utilize the post-hoc setting in \cite{panthaplackel2021deep}, where the code changes that resulted in the mismatch are unknown; Only the current version of the code snippet and the old comment are available. This setting is similar to our work, where we want to detect inconsistency for code-text pairs.

\paragraph{Code Summarization:} This task aims to generate a natural language summary from a given piece of code, playing a significant role in the domain of software comprehension and documentation. 
By summarizing key concepts and features into a concise format, code summarization addresses the challenge of comprehending programming constructs, especially as codebases continue to grow in complexity.

\subsection{Datasets} 
As we assess the performance of DocChecker across two distinct tasks, we rely on two datasets: the Just-In-Time dataset for the ICCD task and the CodeXGLUE dataset for the code summarization task.

\paragraph{Just-In-Time Dataset:}
In this dataset, each sample is the comment-method pair from 2 versions: before and after updating $(C_1, M_1)$ and $(C_2, M_2)$ \cite{panthaplackel2021deep}. In the post-hoc setting, $C=C_1$ and $M=M_2$. They assume that the developer updated the comment because it became inconsistent as a result of code changes; they take $C_1$ to be inconsistent with $M_2$, consequently leading to a negative example. For positive examples, they additionally examine cases in which $C_1=C_2$ and assume that the existing comment has been revised to align consistently with the corresponding code snippet.
For a more reliable evaluation, they manually check to get 300 clean examples from the test set and note it as the cleaned test set.

\paragraph{CodeXGLUE dataset:} 
This dataset comprises six programming languages: Python, Java, JavaScript, Ruby, Go, and PHP. They come from publicly available open source non-fork GitHub repositories, with each documentation representing the first paragraph.  \citet{lu2021codexglue} employed rule-based methods to remove the redundant information and enhance the overall quality of the dataset compared to its original version from \citet{husain2019codesearchnet}.


\subsection{Baselines:} 
\paragraph{Baselines for ICCD task:} We select the following existing work to compare against \methodname for its effectiveness on the ICCD task:
\begin{itemize}[leftmargin=*]
    \item 
 \textbf{SVM \cite{corazza2018coherence}}: This bag-of-words approach classifies whether a comment is coherent with the method using an SVM with TF-IDF vectors corresponding to the comment and method; 
\item  Since we first pre-train \methodname and fine-tune it for downstream tasks, we also compare our method to two effective pre-trained models: \textbf{CodeBERT} \cite{feng2020codebert} and \textbf{CodeT5} \cite{wang2021codet5}.
\item \textbf{Deep-JIT}: \cite{panthaplackel2021deep} presents a method for detecting inconsistencies between natural language comments and source code. With different ways of encoding the method, they consider three types and note them as \textit{SEQ, GRAPH, HYBRID}. Deep-JIT is the existing SOTA method on the Just-In-Time dataset. 

\item CodeLLMs: We choose three widely-used and powerful CodeLLMs: \textbf{GPT-3.5-Turbo}, \textbf{StarCoder} (15B) \cite{li2023starcoder} and \textbf{CodeLlama} (34B) \cite{roziere2023code}. We evaluate LLMs in this task using zero-shot (0-shot) and few-shot (3-shot) prompting. In zero-shot prompts, no examples from the Just-In-Time dataset are provided, while in the few-shot experiment, each prompt includes three code-text pairs along with the correct label from the dataset. Both sets of prompts are run on all selected LLMs.
\end{itemize}

\paragraph{Baselines for Code Summarization task:}  In this experiment, we focus on the fine-tuning setting and compare our method with smaller-scale LMs, including RoBERTa \cite{liu2019roberta}, CodeBERT \cite{feng2020codebert} trained with masked language modeling; PLBART \cite{ahmad2021unified} is based on BART and pre-trained using denoising objective; CodeT5 \cite{wang2021codet5}, adapted from T5, takes into account important token-type information in identifiers; and the variant of UniXcoder \cite{guo2022unixcoder} since we utilize UniXcoder as the backbone network.

\subsection{Metrics}
\paragraph{Metrics for ICCD:} We use two common classification metrics: F1 score (w.r.t. the positive label) and Accuracy (Acc) to report the performance of the methods.

\paragraph{Metrics for Code Summarization:} For this task, we use the smoothed BLEU-4 \cite{lin2004orange} as the evaluation metric and report the overall score of six programming languages.

\begin{table}[]
	\centering
        \small
    \begin{tabular}{@{}cccccc@{}}
\toprule
\multirow{2}{*}{Method}                      & \multicolumn{2}{c}{Cleaned Test set}                                                                    &                      & \multicolumn{2}{c}{Full Test Set}                                                                       \\ \cmidrule(l){2-6} 
                                             & F1                                                 & Acc                                                &                      & F1                                                 & Acc                                                \\ \midrule
SVM & 53.9                                               & 60.7                                               &                      & 54.6                                               & 60.3                                               \\
$\text{Deep-JIT}_\text{SEQ}$             & 63.0                                               & 60.3                                               &                      & 66.3                                               & 62.8                                               \\
$\text{Deep-JIT}_\text{GRAPH}$           & 65.0                                               & 62.2                                               &                      & 67.2                                               & 64.6                                               \\
$\text{Deep-JIT}_\text{HYBRID}$          & 63.3                                               & 55.2                                              &                      & 66.3                                               & 58.9                                               \\ \cmidrule(r){1-3} \cmidrule(l){5-6} 
CodeBERT                                     & 67.9                                               & 66.9                                               &                      & 70.7                                               & 69.8                                               \\
CodeT5                                       & 69.5                                               & 68.8                                               &                      & 70.2                                               & 70.1                                                \\ 

\cmidrule(r){1-3} \cmidrule(l){5-6} 
$\text{GPT-3.5 }_\text{0-shot}$                                      & 60.9                                               &  65.1                                              &                     &    62.5                                            & 64.6                                                \\ 
$\text{StarCoder }_\text{0-shot}$                                        & 43.7                                               & 43.1                                               &                      & 45.2                                               & 43.9                                              \\ 
$\text{CodeLlama }_\text{0-shot}$                                          & 70.2                                               & 68.7                                              &                      & 62.6                                               & 61.8                                              \\ 
\cmidrule(r){1-3} \cmidrule(l){5-6} 
$\text{GPT-3.5 }_\text{3-shot}$                                      & 66.4                                               & 67.0                                               &                     & 66.1                                               & 61.4                                                \\ 
$\text{StarCoder }_\text{3-shot}$                                       & 44.2                                               & 43.6                                               &                      & 42.8                                               & 42.2                                              \\ 
$\text{CodeLlama }_\text{3-shot}$                                          & 70.5                                               & 69.2                                              &                      & 62.3                                               & 62.1                                              \\ \cmidrule(r){1-3} \cmidrule(l){5-6} 
DocChecker                                    &\textbf{ 73.1} & \textbf{70.7} &  & \textbf{74.3} &\textbf{ 72.3} \\ \bottomrule
\end{tabular}
    \caption{Results for post hoc settings on the Just-In-Time dataset}
    \label{tab:just-in-time}
\end{table}

\section{Evaluation Results}
    \subsection{The effectiveness of DocChecker on ICCD}
Table \ref{tab:just-in-time} presents results for all baselines under the post-hoc setting and LLMs. In general, we find that our model can significantly outperform all of the baselines. 
Despite CodeBERT and CodeT5 being pre-trained models with more parameters, showcasing efficiency in numerous downstream tasks, their performance is behind ours. DocChecker achieves a new SoTA of 72.3\% accuracy and 74.3\% F1 score on the full test set of Just-In-Time.

On the other hand, although previous literature has empirically explored various capabilities of LLMs in diverse natural language processing and code generation tasks, billion-parameter LLMs such as StarCoder, GPT 3.5, and CodeLlama still struggle with ICCD, even with the construction of various types of prompts. In particular, DocChecker produces significant improvements of +10\% accuracy and F1 score compared to the selected LLMs.

The experiment results suggest that \methodname benefits from using a pre-trained deep learning model with the three above objectives. It supports that our method effectively detects inconsistent samples in the code corpus.


\subsection{The effectiveness of DocChecker on Code Summarization}

\begin{table}[]
\centering
\small
\begin{tabular}{lc}
\hline
\multirow{2}{*}{Method} & Summarization  \\ \cline{2-2} 
                        & BLEU-4         \\ \hline
RoBERTa                 & 16.57          \\
CodeBERT                & 17.83          \\ \hline
PLBART                  & 18.32          \\
CodeT5-small            & 19.14          \\
CodeT5-base             & 19.55          \\ \hline
UniXcoder               & 19.30          \\
-w/o contras            & 19.20          \\
-w/o cross-gen          & 19.27          \\
-w/o comment            & 18.97          \\
-w/o AST                & 19.33          \\
-using BFS              & 19.24          \\
-using DFS              & 19.25          \\ \hline
DocChecker              & \textbf{33.64} \\ \hline
\end{tabular}
\caption{Results on the code summarization task.}
\label{tab:code_summaarization}
\end{table}

Our results on this task are shown in Table \ref{tab:code_summaarization}.
DocChecker is compared to a number of pre-trained code language models in our evaluation. Following DocChecker's pre-training with the three aforementioned objectives, our method outperforms others significantly. DocChecker's BLEU-4 score is twice as high as that of RoBERTa and CodeBERT. 
Furthermore, though CodeT5-base employs a 12-layer encoder and a 12-layer decoder, which is two-fold the capacity of our architecture, its result is significantly lower than ours. DocChecker yields substantial improvements over CodeT5 and the backbone network UniXcoder by +13 BLEU-4 score.

\subsection{The influence of the backbone network on DocChecker}
\label{effective_backbone}

As DocChecker functions as a framework, the selection of an encoder-decoder model for the backbone network is flexible. In this section, we illustrate the impact of several pre-trained models on DocChecker's effectiveness.
We integrate CodeBERT, CodeT5, and UniXcoder as the backbone network for DocChecker. Each chosen backbone undergoes pre-training within the DocChecker framework and is fine-tuned on the Just-In-Time dataset. The results, presented in Table \ref{tab:different_backbone}, reveal that the pre-trained models exhibit improved performance after re-pre-training compared to their original versions. However, UniXcoder emerges as the most effective backbone model for this task, leading us to adopt it for all our experiments.

\begin{table}[]
\centering
\small
\begin{tabular}{lcclcc}
\hline
\multirow{2}{*}{\begin{tabular}[c]{@{}l@{}}Backbone \\ Network\end{tabular}} & \multicolumn{2}{c}{Cleaned test set} &  & \multicolumn{2}{c}{Full test set} \\ \cline{2-6} 
                                                                             & F1                & Acc              &  & F1              & Acc             \\ \hline
CodeBERT                                                                     & 68.2              & 67.1             &  & 71.5            & 70.4            \\
CodeT5                                                                       & 70.1              & 69.5             &  & 71.9            & 71.5            \\
UniXcoder                                                                    & 73.1              & 70.7             &  & 74.3            & 72.3            \\ \hline
\end{tabular}
\caption{Results of DocChecker pre-trained with different backbone networks on the Just-In-Time dataset.}
\label{tab:different_backbone}
\end{table}

\subsection{Practical Application}

Besides showing DocChecker's performance, we highlight its efficacy in real-world scenarios. 
We consider the widely used CodeXGlue dataset, which extracts functions and their paired comments from Github repositories.
Although this benchmark dataset is expected to be of good quality, noise is inevitable due to the differences in coding conventions and assumptions employed in modern programming languages and IDEs.
Using DocChecker, we can filter the inconsistent code-comment samples within the dataset and generate new comprehensive summary sentences for them.

\begin{figure}[]
\centering
\tiny
\includegraphics[width=0.4\textwidth]{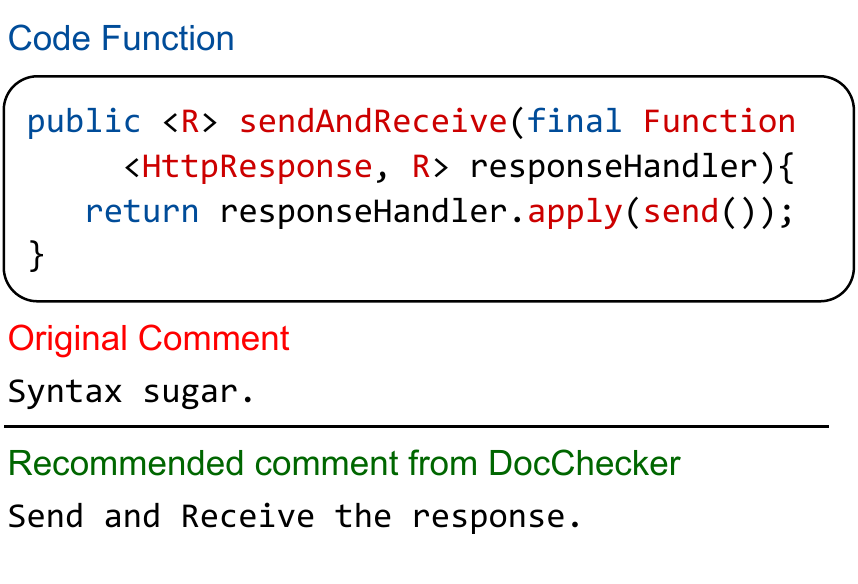}
\caption{An inconsistent code-comment example collected from the CodeXGlue dataset and our recommended docstring to replace.
}
\label{fig:noise_samples}
\end{figure}

Figure \ref{fig:noise_samples} illustrates an example of an inconsistent sample detected by \methodname in the CodeSearchNet dataset. The comment associated with the code snippet is misaligned and requires an update. Beyond detection, our method also provides a detailed summary sentence for each sample, serving as a replacement for the outdated ones.

\section{Conclusion}
    In this paper, we present DocChecker, a tool to filter and generate replacement comments for inconsistent code-comment pairs. The experimental results demonstrate the effectiveness of this method compared to the SoTA existing methods and LLMs, showcasing its applicability in both academic and practical contexts. We have released \methodname as an easy-to-use library, complemented by a user-friendly interface to enhance user interaction.

\bibliography{anthology,reference}

\begin{thebibliography}{32}
\expandafter\ifx\csname natexlab\endcsname\relax\def\natexlab#1{#1}\fi

\bibitem[{Ahmad et~al.(2021)Ahmad, Chakraborty, Ray, and
  Chang}]{ahmad2021unified}
Wasi~Uddin Ahmad, Saikat Chakraborty, Baishakhi Ray, and Kai-Wei Chang. 2021.
\newblock Unified pre-training for program understanding and generation.
\newblock \emph{arXiv preprint arXiv:2103.06333}.

\bibitem[{Corazza et~al.(2018)Corazza, Maggio, and
  Scanniello}]{corazza2018coherence}
Anna Corazza, Valerio Maggio, and Giuseppe Scanniello. 2018.
\newblock Coherence of comments and method implementations: a dataset and an
  empirical investigation.
\newblock \emph{Software Quality Journal}, 26:751--777.

\bibitem[{Devlin et~al.(2019)Devlin, Chang, Lee, and
  Toutanova}]{DBLP:conf/naacl/DevlinCLT19}
Jacob Devlin, Ming{-}Wei Chang, Kenton Lee, and Kristina Toutanova. 2019.
\newblock {BERT:} pre-training of deep bidirectional transformers for language
  understanding.
\newblock In \emph{Proceedings of the 2019 Conference of the North American
  Chapter of the Association for Computational Linguistics: Human Language
  Technologies, {NAACL-HLT} 2019, Minneapolis, MN, USA, June 2-7, 2019, Volume
  1 (Long and Short Papers)}, pages 4171--4186. Association for Computational
  Linguistics.

\bibitem[{Di~Grazia and Pradel(2023)}]{codesearchsurvey}
Luca Di~Grazia and Michael Pradel. 2023.
\newblock Code search: A survey of techniques for finding code.
\newblock \emph{ACM Comput. Surv.}, 55(11).

\bibitem[{Feng et~al.(2020)Feng, Guo, Tang, Duan, Feng, Gong, Shou, Qin, Liu,
  Jiang, and Zhou}]{feng2020codebert}
Zhangyin Feng, Daya Guo, Duyu Tang, Nan Duan, Xiaocheng Feng, Ming Gong, Linjun
  Shou, Bing Qin, Ting Liu, Daxin Jiang, and Ming Zhou. 2020.
\newblock \href {http://arxiv.org/abs/2002.08155} {Codebert: A pre-trained
  model for programming and natural languages}.

\bibitem[{Guo et~al.(2022)Guo, Lu, Duan, Wang, Zhou, and
  Yin}]{guo2022unixcoder}
Daya Guo, Shuai Lu, Nan Duan, Yanlin Wang, Ming Zhou, and Jian Yin. 2022.
\newblock \href {https://doi.org/10.18653/v1/2022.acl-long.499} {Unixcoder:
  Unified cross-modal pre-training for code representation}.
\newblock pages 7212--7225.

\bibitem[{Husain et~al.(2019)Husain, Wu, Gazit, Allamanis, and
  Brockschmidt}]{husain2019codesearchnet}
Hamel Husain, Ho{-}Hsiang Wu, Tiferet Gazit, Miltiadis Allamanis, and Marc
  Brockschmidt. 2019.
\newblock \href {http://arxiv.org/abs/1909.09436} {Codesearchnet challenge:
  Evaluating the state of semantic code search}.
\newblock \emph{CoRR}, abs/1909.09436.

\bibitem[{Kanade et~al.(2020)Kanade, Maniatis, Balakrishnan, and Shi}]{cubert}
Aditya Kanade, Petros Maniatis, Gogul Balakrishnan, and Kensen Shi. 2020.
\newblock Learning and evaluating contextual embedding of source code.
\newblock In \emph{Proceedings of the 37th International Conference on Machine
  Learning}, ICML'20. JMLR.org.

\bibitem[{Li et~al.(2021)Li, Selvaraju, Gotmare, Joty, Xiong, and
  Hoi}]{li2021align}
Junnan Li, Ramprasaath Selvaraju, Akhilesh Gotmare, Shafiq Joty, Caiming Xiong,
  and Steven Chu~Hong Hoi. 2021.
\newblock Align before fuse: Vision and language representation learning with
  momentum distillation.
\newblock \emph{Advances in neural information processing systems},
  34:9694--9705.

\bibitem[{Li et~al.(2023)Li, Allal, Zi, Muennighoff, Kocetkov, Mou, Marone,
  Akiki, Li, Chim et~al.}]{li2023starcoder}
Raymond Li, Loubna~Ben Allal, Yangtian Zi, Niklas Muennighoff, Denis Kocetkov,
  Chenghao Mou, Marc Marone, Christopher Akiki, Jia Li, Jenny Chim, et~al.
  2023.
\newblock Starcoder: may the source be with you!
\newblock \emph{arXiv preprint arXiv:2305.06161}.

\bibitem[{Lin et~al.(2021)Lin, Wang, Liu, Mao, and Bissyandé}]{hebCUP}
Bo~Lin, Shangwen Wang, Kui Liu, Xiaoguang Mao, and Tegawendé~F. Bissyandé.
  2021.
\newblock \href {https://doi.org/10.1109/ICPC52881.2021.00013} {Automated
  comment update: How far are we?}
\newblock In \emph{2021 IEEE/ACM 29th International Conference on Program
  Comprehension (ICPC)}, pages 36--46.

\bibitem[{Lin and Och(2004)}]{lin2004orange}
Chin-Yew Lin and Franz~Josef Och. 2004.
\newblock Orange: a method for evaluating automatic evaluation metrics for
  machine translation.
\newblock In \emph{COLING 2004: Proceedings of the 20th International
  Conference on Computational Linguistics}, pages 501--507.

\bibitem[{Liu et~al.(2019)Liu, Ott, Goyal, Du, Joshi, Chen, Levy, Lewis,
  Zettlemoyer, and Stoyanov}]{liu2019roberta}
Yinhan Liu, Myle Ott, Naman Goyal, Jingfei Du, Mandar Joshi, Danqi Chen, Omer
  Levy, Mike Lewis, Luke Zettlemoyer, and Veselin Stoyanov. 2019.
\newblock \href {http://arxiv.org/abs/1907.11692} {Roberta: {A} robustly
  optimized {BERT} pretraining approach}.
\newblock \emph{CoRR}, abs/1907.11692.

\bibitem[{Liu et~al.(2021)Liu, Xia, Lo, Yan, and Li}]{CUP}
Zhongxin Liu, Xin Xia, David Lo, Meng Yan, and Shanping Li. 2021.
\newblock Just-in-time obsolete comment detection and update.
\newblock \emph{IEEE Transactions on Software Engineering}, pages 1--1.

\bibitem[{Lu et~al.(2021)Lu, Guo, Ren, Huang, Svyatkovskiy, Blanco, Clement,
  Drain, Jiang, Tang, Li, Zhou, Shou, Zhou, Tufano, GONG, Zhou, Duan,
  Sundaresan, Deng, Fu, and LIU}]{lu2021codexglue}
Shuai Lu, Daya Guo, Shuo Ren, Junjie Huang, Alexey Svyatkovskiy, Ambrosio
  Blanco, Colin Clement, Dawn Drain, Daxin Jiang, Duyu Tang, Ge~Li, Lidong
  Zhou, Linjun Shou, Long Zhou, Michele Tufano, MING GONG, Ming Zhou, Nan Duan,
  Neel Sundaresan, Shao~Kun Deng, Shengyu Fu, and Shujie LIU. 2021.
\newblock \href {https://openreview.net/forum?id=6lE4dQXaUcb} {Code{XGLUE}: A
  machine learning benchmark dataset for code understanding and generation}.
\newblock In \emph{Thirty-fifth Conference on Neural Information Processing
  Systems Datasets and Benchmarks Track (Round 1)}.

\bibitem[{Manh et~al.(2023)Manh, Hai, Dau, Nguyen, Nghiem, Guo, and
  Bui}]{manh2023vault}
Dung~Nguyen Manh, Nam~Le Hai, Anh T.~V. Dau, Anh~Minh Nguyen, Khanh Nghiem, Jin
  Guo, and Nghi D.~Q. Bui. 2023.
\newblock \href {http://arxiv.org/abs/2305.06156} {The vault: A comprehensive
  multilingual dataset for advancing code understanding and generation}.

\bibitem[{Nijkamp et~al.(2022)Nijkamp, Pang, Hayashi, Tu, Wang, Zhou, Savarese,
  and Xiong}]{nijkamp2022codegen}
Erik Nijkamp, Bo~Pang, Hiroaki Hayashi, Lifu Tu, Huan Wang, Yingbo Zhou, Silvio
  Savarese, and Caiming Xiong. 2022.
\newblock Codegen: An open large language model for code with multi-turn
  program synthesis.
\newblock \emph{arXiv preprint arXiv:2203.13474}.

\bibitem[{Panthaplackel et~al.(2021)Panthaplackel, Li, Gligoric, and
  Mooney}]{panthaplackel2021deep}
Sheena Panthaplackel, Junyi~Jessy Li, Milos Gligoric, and Raymond~J Mooney.
  2021.
\newblock Deep just-in-time inconsistency detection between comments and source
  code.
\newblock In \emph{Proceedings of the AAAI Conference on Artificial
  Intelligence}, volume~35, pages 427--435.

\bibitem[{Rabbi and Siddik(2020)}]{detecting}
Fazle Rabbi and Md~Saeed Siddik. 2020.
\newblock Detecting code comment inconsistency using siamese recurrent network.
\newblock In \emph{Proceedings of the 28th International Conference on Program
  Comprehension}, ICPC '20, page 371–375. Association for Computing
  Machinery.

\bibitem[{Radford et~al.(2018)Radford, Narasimhan, Salimans, Sutskever
  et~al.}]{radford2018improving}
Alec Radford, Karthik Narasimhan, Tim Salimans, Ilya Sutskever, et~al. 2018.
\newblock Improving language understanding by generative pre-training.

\bibitem[{Ratol and Robillard(2017)}]{ratol2017detecting}
Inderjot~Kaur Ratol and Martin~P Robillard. 2017.
\newblock Detecting fragile comments.
\newblock In \emph{2017 32nd IEEE/ACM International Conference on Automated
  Software Engineering (ASE)}, pages 112--122. IEEE.

\bibitem[{Roziere et~al.(2023)Roziere, Gehring, Gloeckle, Sootla, Gat, Tan,
  Adi, Liu, Remez, Rapin et~al.}]{roziere2023code}
Baptiste Roziere, Jonas Gehring, Fabian Gloeckle, Sten Sootla, Itai Gat,
  Xiaoqing~Ellen Tan, Yossi Adi, Jingyu Liu, Tal Remez, J{\'e}r{\'e}my Rapin,
  et~al. 2023.
\newblock Code llama: Open foundation models for code.
\newblock \emph{arXiv preprint arXiv:2308.12950}.

\bibitem[{Shi et~al.(2022)Shi, Mu, Chen, Wang, Wang, Yang, Li, Xia, and
  Wang}]{shi2022we}
Lin Shi, Fangwen Mu, Xiao Chen, Song Wang, Junjie Wang, Ye~Yang, Ge~Li, Xin
  Xia, and Qing Wang. 2022.
\newblock Are we building on the rock? on the importance of data preprocessing
  for code summarization.
\newblock In \emph{Proceedings of the 30th ACM Joint European Software
  Engineering Conference and Symposium on the Foundations of Software
  Engineering}, pages 107--119.

\bibitem[{Steiner and Zhang(2022)}]{steiner2022code}
Theo Steiner and Rui Zhang. 2022.
\newblock Code comment inconsistency detection with bert and longformer.
\newblock \emph{arXiv preprint arXiv:2207.14444}.

\bibitem[{Sun et~al.(2022)Sun, Li, Liu, Du, and Li}]{sun2022importance}
Zhensu Sun, Li~Li, Yan Liu, Xiaoning Du, and Li~Li. 2022.
\newblock On the importance of building high-quality training datasets for
  neural code search.
\newblock In \emph{Proceedings of the 44th International Conference on Software
  Engineering}, pages 1609--1620.

\bibitem[{Svyatkovskiy et~al.(2020)Svyatkovskiy, Deng, Fu, and
  Sundaresan}]{IntelliCode}
Alexey Svyatkovskiy, Shao~Kun Deng, Shengyu Fu, and Neel Sundaresan. 2020.
\newblock Intellicode compose: Code generation using transformer.
\newblock ESEC/FSE 2020, page 1433–1443.

\bibitem[{Tan et~al.(2012)Tan, Marinov, Tan, and Leavens}]{6200082}
Shin~Hwei Tan, Darko Marinov, Lin Tan, and Gary~T. Leavens. 2012.
\newblock \href {https://doi.org/10.1109/ICST.2012.106} {@tcomment: Testing
  javadoc comments to detect comment-code inconsistencies}.
\newblock In \emph{2012 IEEE Fifth International Conference on Software
  Testing, Verification and Validation}, pages 260--269.

\bibitem[{Vaswani et~al.(2017)Vaswani, Shazeer, Parmar, Uszkoreit, Jones,
  Gomez, Kaiser, and Polosukhin}]{vaswani2017attention}
Ashish Vaswani, Noam Shazeer, Niki Parmar, Jakob Uszkoreit, Llion Jones,
  Aidan~N Gomez, {\L}ukasz Kaiser, and Illia Polosukhin. 2017.
\newblock Attention is all you need.
\newblock \emph{Advances in neural information processing systems}, 30.

\bibitem[{Wang et~al.(2023)Wang, Le, Gotmare, Bui, Li, and
  Hoi}]{wang2023codet5+}
Yue Wang, Hung Le, Akhilesh~Deepak Gotmare, Nghi~DQ Bui, Junnan Li, and
  Steven~CH Hoi. 2023.
\newblock Codet5+: Open code large language models for code understanding and
  generation.
\newblock \emph{arXiv preprint arXiv:2305.07922}.

\bibitem[{Wang et~al.(2021)Wang, Wang, Joty, and Hoi}]{wang2021codet5}
Yue Wang, Weishi Wang, Shafiq Joty, and Steven C.~H. Hoi. 2021.
\newblock \href {http://arxiv.org/abs/2109.00859} {Codet5: Identifier-aware
  unified pre-trained encoder-decoder models for code understanding and
  generation}.

\bibitem[{Wen et~al.(2019)Wen, Nagy, Bavota, and Lanza}]{wen2019large}
Fengcai Wen, Csaba Nagy, Gabriele Bavota, and Michele Lanza. 2019.
\newblock A large-scale empirical study on code-comment inconsistencies.
\newblock In \emph{2019 IEEE/ACM 27th International Conference on Program
  Comprehension (ICPC)}, pages 53--64. IEEE.

\bibitem[{Xu et~al.(2022)Xu, Alon, Neubig, and Hellendoorn}]{codeX}
Frank~F. Xu, Uri Alon, Graham Neubig, and Vincent~Josua Hellendoorn. 2022.
\newblock \href {https://doi.org/10.1145/3520312.3534862} {A systematic
  evaluation of large language models of code}.
\newblock In \emph{Proceedings of the 6th ACM SIGPLAN International Symposium
  on Machine Programming}, MAPS 2022, page 1–10, New York, NY, USA.
  Association for Computing Machinery.

\end{thebibliography}

\appendix


\end{document}